\documentclass[aps,prl,preprint,superscriptaddress]{revtex4-1}
\usepackage{graphicx}
\usepackage{dcolumn}
\usepackage{bm}
\usepackage[latin1]{inputenc}
\usepackage{color}
\begin{document}


\title{Role of the radiation-reaction electric field in the optical response of two-dimensional crystals}


\author{Michele Merano}
\email[]{michele.merano@unipd.it}
\affiliation{Dipartimento di Fisica e Astronomia G. Galilei, Universit$\grave{a}$ degli studi di Padova, via Marzolo 8, 35131 Padova, Italy}


\date{\today}

\begin{abstract}
A classical theory of a radiating two-dimensional crystal is proposed and an expression for the radiative-reaction electric field is derived. This field plays an essential role in connecting the microscopic electromagnetic fields acting on each dipole of the crystal to the macroscopic one, via the boundary conditions for the system. The expression of the radiative-reaction electric field coincides with the macroscopic electric field radiating from the crystal and, summed to the incident electric field, generates the total macroscopic electric field.
\end{abstract}


\maketitle

\section{Introduction}

The discovery of graphene and other two-dimensional (2D) crystals in 2004 was a major scientific breakthrough in material science \cite{Novoselov2004, Novoselov2005}. In the past decade the family of 2D crystals has widened encompassing a large selection of compositions including almost all the elements of the periodic table \cite{Heine14}. The variety of their electronic properties is extremely broad, including metals and semi-metals (for example graphene \cite{Novoselov2004, CastroNeto09}), insulators (hexagonal boron nitride \cite{Blake2011}) and semiconductors (transition-metal dichalcogenides \cite{Heinz2010, Heinz2014}). These single-layer atomic crystals are stable under ambient conditions and appear continuous on a macroscopic scale  \cite{Novoselov2005}. 

Macroscopic continuity is confirmed also by their optical properties \cite{Ramasubramaniam14, Nair2008, Blake2007, Heinz2014, Blake2011, Kravets2010}. In analogy with 3D materials where the susceptibility and the conductivity fix the optical response, single-layer atomic crystals are described as zero-thickness sheets with a surface susceptibility and a surface conductivity \cite{Pershoguba07, Hanson08, Merano15, Merano16}. As for bulk materials, ellipsometry is able to furnish both of these parameters, showing that these are the macroscopic physical quantities experimentally accessible from the linear optical response of 2D atomic crystals \cite{Merano16}. Also the nonlinear optical properties are described by modelling them as zero-thickness sheets with a nonlinear surface susceptibility \cite{Zhao13, Heinz13, Merano216}.      

At the microscopic level, in analogy to what happens for a bulk crystal, it is possible to define a local electric field that connects the macroscopic surface electric susceptibility to the atomic polarizability \cite{Luca16}. This classical approach provides a deep insight. It explains why the Fresnel coefficients for a single-layer 2D atomic crystal are intrinsically complex quantities even when a null surface conductivity is assumed. The crystal is modelled as atoms with polarizability $\alpha$ distributed on a Bravais lattice. Contrary to the 3D case \cite{Aspnes82, Wolf} all of the dipoles contribute to the local electric field and the retardation effects (due to the finite velocity of propagation of the electromagnetic potentials) dephase the local electric field, and hence the macroscopic surface polarization, from the electric field incident on the crystal \cite{Luca16}.

Here this classical approach is further extended to treat in a complete way the radiating properties of a 2D crystal. Improvements to the microscopic description must include two directions that in the end will merge. First, the radiation-reaction electric field acting on each dipole needs to be computed. In theories dealing with the local electric field for 2D crystals \cite{Luca16} and metasurfaces \cite{Kuester03} only the retarded potentials have been considered, while radiation-reaction processes are better taken into account by addressing also the role of the advanced potentials \cite{Page24, Dirac38, Wheeler45, Panofsky}. Second this microscopic description must be connected to the macroscopic one. The local electric field, acting on each dipole, is different from the macroscopic field in the 2D crystal. Anyway a complete theory should be able to provide also this last one \cite{Kuester03}. 

\section{The mechanism of the radiative reaction}

The origin of the force of radiative reaction has puzzled scientists for a long time. Use of action at a distance with field theory as equivalent and complementary tools for the description of nature has for long been prevented by inability of the first point of view fully to account for the mechanism of radiation. Elucidation of this point came through the combined effort of many physicists, including Lorentz \cite{Panofsky}, Page \cite{Page24}, Dirac \cite{Dirac38}, Wheeler and Feynman \cite{Wheeler45}. 

For the purpose of this paper it is interesting to refer back to the calculation of radiative reaction made by Lorentz on an extended charge, every part of which exerted a retarded effect upon every other part \cite{Panofsky}. This calculation was further examined in \cite{Page24, Dirac38, Wheeler45} to produce a well-defined and relativistically invariant prescription to compute the magnitude of the radiative damping. If we consider a system of $n$ charged particles, the electromagnetic fields which act on a given particle $j$ arise only from other particles \cite{Wheeler45}. These fields are represented by one-half of the retarded plus one-half of the advanced Lienard-Wiechert solutions of Maxwell's equations \cite{Dirac38, Wheeler45, Panofsky}. The radiation-reaction electric field can be defined as the difference between one-half of the retarded fields and one-half of the advanced fields generated by the other particles \cite{Page24, Dirac38, Wheeler45}. This field compensates the half-advanced fields and combines with its half-retarded fields, to produce the full retarded disturbance $\vec{\textbf{\emph{E}}}_{j}$ which is required by experience \cite{Wheeler45}  
\begin{eqnarray}
\label{Retarded field}
\vec{\textbf{\emph{E}}}_{j}&=&\frac{1}{2}{\sum_{n}}'\bigg(\vec{\textbf{\emph{E}}}_{ret, n}+\vec{\textbf{\emph{E}}}_{adv, n}\bigg) 
+\frac{1}{2}{\sum_{n}}'\bigg(\vec{\textbf{\emph{E}}}_{ret, n}-\vec{\textbf{\emph{E}}}_{adv, n}\bigg)
\end{eqnarray}
where the dash in the summation indicates that the $j$ particle is excluded from the sum. This procedure is not convenient to apply to cases where relativistic contraction is so strong so as to modify the form of the crystal \cite{Wheeler45}, such cases will not be considered here.

\section{Radiation-reaction electric field in 2D crystals}

A 2D crystal can be treated as an extended charge distribution or better an extended electric dipoles distribution \cite{Luca16}.  I will find an expression for the radiation-reaction electric field $\vec{\textbf{\emph{E}}}_{R}$ using the above prescription that is well-defined and involves no more than standard electromagnetic theory. An argument based on conservation of energy will further confirm that the proposed expression is the correct one \cite{Jackson}. Finally, $\vec{\textbf{\emph{E}}}_{R}$ will be connected with the macroscopic electric field. 

\begin{figure}
\includegraphics{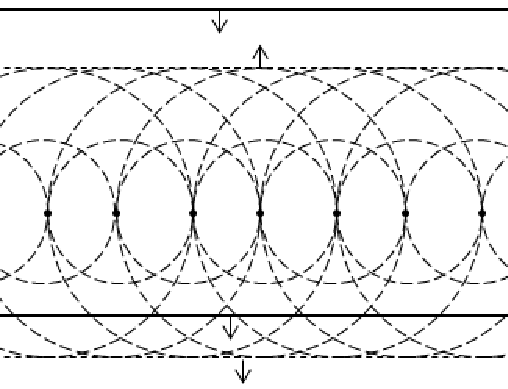}
\caption{The spherical waves depicted here represent light reemitted by the dipoles (black dots) in a 2D crystal that is exposed to a plane wave entering from the top (solid line). The reflected wave is traveling to the top. The transmitted wave is a coherent superposition of the spherical wavelets and the incident wave.}
\end{figure}

\subsection{Classical model of radiating 2D crystals}

I consider an insulating 2D crystal formed by atoms with polarizability $\alpha$ placed on a 2D Bravais lattice. A linearly polarized electromagnetic plane wave is incident on this 2D crystal (Fig.1). For simplicity normal incidence is assumed. The local electric field $\vec{\textbf{\emph{E}}}_{loc}$ acting on a single dipole (for instance the one in the origin) is given by
\begin{eqnarray}
\label{Local2}
\vec{\textbf{\emph{E}}}_{loc}e^{i(\omega t + \varphi)}=\vec{\textbf{\emph{E}}}_{i}e^{i\omega t}
+{\sum_{(m,n)}}'\vec{\textbf{\emph{E}}}_{n,m}(t)
\end{eqnarray}
where $\vec{\textbf{\emph{E}}}_{loc}$ and $\vec{\textbf{\emph{E}}}_{i}$ are real vectors, $\vec{\textbf{\emph{E}}}_{i}$ is the incident electric field and the last term is the full retarded disturbance which is required by experience \cite{Wheeler45}. This last term was computed in \cite{Luca16} for three different dipole distributions: square, triangular and honeycomb lattices. Formula (\ref{Local2}) shows that the dipole fields dephase the local field with respect to the incident electric field. On the macroscopic scale this translates to a dephasing between the incident electric field and the surface polarization density $\vec{\textbf{\emph{P}}}(t)$
\begin{eqnarray}
\label{Local1}
\vec{\textbf{\emph{P}}}(t)=N\vec{\textbf{\emph{p}}}(t)=N\alpha \epsilon_{0} \vec{\textbf{\emph{E}}}_{loc}e^{i(\omega t + \varphi)}
\end{eqnarray}
where $\vec{\textbf{\emph{p}}}$ is the induced dipole moment at each reticular point, $N$ is the number of primitive cells per unit area and $\epsilon_{0}$ is the vacuum permittivity. 

\subsection{Advanced and retarded electric fields for a dipole}

The theory developed in \cite{Luca16} used only retarded fields. As explained above a complete treatment of the radiating crystal requires also the advanced fields. Following ref. \cite{Feynman} the field generated by an electric dipole is   
\begin{eqnarray}
\label{Dipole2}
\vec{\textbf{\emph{E}}}(\vec{\textbf{\emph{r}}}, t)=\frac{1}{4\pi \epsilon_{0}r^3} \biggl(3(\tilde{\vec{\textbf{\emph{p}}}}\cdot\hat{\textbf{\emph{r}}})\hat{\textbf{\emph{r}}}-\tilde{\vec{\textbf{\emph{p}}}} -\frac{( \vec{\textbf{\emph{r}}} \times \ddot{\vec{\textbf{\emph{p}}}})  \times \vec{\textbf{\emph{r}}}}{c^2}  \biggr)
\end{eqnarray}
where 
\begin{eqnarray}
\vec{\textbf{\emph{p}}}&=&\vec{\textbf{\emph{p}}}(t \pm \frac{r}{c}) \nonumber \\
\tilde{\vec{\textbf{\emph{p}}}}&=&\vec{\textbf{\emph{p}}}(t \pm \frac{r}{c}) \mp \frac{r}{c}\dot{\vec{\textbf{\emph{p}}}}(t \pm \frac{r}{c})
\end{eqnarray}
here the $\pm$ sign in parentheses means that the dipole moment must be evaluated at the corresponding advanced or retarded time.
Assuming a temporal dependence $e^{i \omega t}$
\begin{eqnarray}
\vec{\textbf{\emph{p}}}=\vec{\textbf{\emph{p}}}_0 e^{i(\omega t \pm kr)}
\end{eqnarray}
I obtain:
\begin{eqnarray}
\label{Dipole3}
\vec{\textbf{\emph{E}}}(\vec{\textbf{\emph{r}}}, t)&=&\frac{e^{i(\omega t \pm kr)}}{4\pi \epsilon_{0}r^3} \biggl((3(\vec{\textbf{\emph{p}}}_0\cdot\hat{\textbf{\emph{r}}})\hat{\textbf{\emph{r}}}-\vec{\textbf{\emph{p}}}_0)(1\mp kr) \nonumber \\
&+&k^2( \vec{\textbf{\emph{r}}}\times \vec{\textbf{\emph{p}}}_0)  \times \vec{\textbf{\emph{r}}}  \biggr)
\end{eqnarray}
As for a plane wave, if the advanced field is 
\begin{eqnarray}
\label{Advanced}
\vec{\textbf{\emph{E}}}_{adv}(\vec{\textbf{\emph{r}}}, t)= e^{i \omega t}\vec{\textbf{\emph{f}}}(\vec{\textbf{\emph{r}}})
\end{eqnarray}
the retarded one is
\begin{eqnarray}
\label{Retarded}
\vec{\textbf{\emph{E}}}_{ret}(\vec{\textbf{\emph{r}}}, t)= e^{i \omega t}\vec{\textbf{\emph{f}}}^*(\vec{\textbf{\emph{r}}})
\end{eqnarray}

\subsection{Expression of the radiation-reaction electric field}

I compute the radiation-reaction electric field acting on a dipole as half the difference between the retarded and the advanced electric fields generated by all the other dipoles. 

\subsubsection{Square and triangular lattices}

For a square and a triangular lattice, the retarded electric fields generated by all the other dipoles is given by (formula (36) of ref. \cite{Luca16})
\begin{eqnarray}
\label{Retarded_square}
{\sum_{(m,n)}}'  \vec{\textbf{\emph{E}}}_{ret, \ m,n}(t)=\frac{\alpha}{4\pi a^3}
\big(C_0+i\, C_1 k a\big)
\vec{\textbf{\emph{E}}}_{loc}(t)
\end{eqnarray} 
where the real and the imaginary part of the summation are proportional respectively to two real constants $C_0$ and $C_1= -2 \pi N a^2$ computed in \cite{Luca16} and $a$ is the lattice spacing.  
From formulas (\ref{Retarded_square}), (\ref{Advanced}) and (\ref{Retarded}) the radiation-reaction electric field acting on a dipole is
\begin{eqnarray}
\label{Reaction_square}
\vec{\textbf{\emph{E}}}_{R}(t) &= &{\sum_{(m,n)}}' \frac{1}{2}\bigg( \vec{\textbf{\emph{E}}}_{ret, \ m,n}(t)-\vec{\textbf{\emph{E}}}_{adv, \ m,n}(t) \bigg) \qquad \nonumber \\ 
&=&\frac{i\, \alpha C_1 k}{4\pi a^2} \vec{\textbf{\emph{E}}}_{loc}(t)=- \frac{\eta N}{2}  \dot{\vec{\textbf{\emph{p}}}}(t)= - \frac{\eta}{2}  \dot{\vec{\textbf{\emph{P}}}}(t)
\end{eqnarray} 
where $\eta$ is the wave impedance of vacuum. 

\subsubsection{Honeycomb lattice}

I consider now the case of a honeycomb lattice with two atoms of different polarizability $\alpha_1$ and $\alpha_2$ per primitive cell. From formula (13) and (43) of ref. \cite{Luca16} the retarded electric fields generated by all the other dipoles is given by
\begin{eqnarray}
\label{Retarded_honey}
{\sum_{(m,n)}}'  \vec{\textbf{\emph{E}}}_{ret, \ m,n}(t)&=&\frac{\alpha_1}{4\pi a^3}
\big(C^{(1)}_0+i\, C_1 k a\big)
\vec{\textbf{\emph{E}}}^{(1)}_{loc}(t) \nonumber \\
&+&\frac{\alpha_2}{4\pi a^3}
\big(C^{(2)}_0+i\, C_1 k a\big)
\vec{\textbf{\emph{E}}}^{(2)}_{loc}(t)
\end{eqnarray} 
where $\vec{\textbf{\emph{E}}}^{(1)}_{loc}(t)$ and $\vec{\textbf{\emph{E}}}^{(2)}_{loc}(t)$ are the local fields acting on the two atoms of the primitive cell, and $C^{(1)}_0$ and $C^{(2)}_0$ are two real constants computed in \cite{Luca16}. From formulas (\ref{Retarded_honey}), (\ref{Advanced}) and (\ref{Retarded}) the radiation-reaction electric field acting on a dipole is 
\begin{eqnarray}
\label{Reaction_honey}
\vec{\textbf{\emph{E}}}_{R}(t) &=& \frac{i\, C_1 k}{4\pi a^2} \bigg(\alpha_1 \vec{\textbf{\emph{E}}}^{(1)}_{loc}(t) + \alpha_2 \vec{\textbf{\emph{E}}}^{(2)}_{loc}(t) \bigg) \nonumber \\ 
&=&- \frac{\eta N}{2} \bigg( \dot{\vec{\textbf{\emph{p}}}}_1(t)+\dot{\vec{\textbf{\emph{p}}}}_2(t)  \bigg)= - \frac{\eta}{2}  \dot{\vec{\textbf{\emph{P}}}}(t)
\end{eqnarray}
Formulas (\ref{Reaction_square}) and (\ref{Reaction_honey}) express the radiation-reaction electric field acting on the single dipoles in terms of their dipole moments. Being proportional to the negative time-derivative dipole moments, this force induces a damping and an energy loss due to the emission of radiation.  

\section{Reflected and transmitted fields from a radiating 2D crystal}

Formulas (\ref{Reaction_square}) and (\ref{Reaction_honey}) are intriguing because they connect a field acting on a single dipole with a macroscopic quantity: the time derivative of the polarization density.  Wherever the polarization in matter changes with time there is an electric current $\vec{\textbf{J}}_{p}$, a genuine motion of charges. The connection between the radiation-reaction electric field and the current density is
\begin{eqnarray}
\label{current density}
\vec{\textbf{J}}_{p}=\dot{\vec{\textbf{\emph{P}}}}(t)= - \frac{2}{\eta}\vec{\textbf{\emph{E}}}_{R}(t)
\end {eqnarray}
This surface current density is a relevant quantity that determines the boundary conditions for the macroscopic fields and hence the Fresnel coefficients of the 2D crystal.

With reference to fig. 2, where a plane wave falls onto the 2D material from the top half space, the boundary conditions for the macroscopic fields are
\begin{eqnarray}
\label{Boundary}
\hat{\kappa} \wedge (\vec{\textbf{E}}_{2}-\vec{\textbf{E}}_{1})=0; \qquad \hat{\kappa} \wedge (\vec{\textbf{H}}_{2}-\vec{\textbf{H}}_{1})=\textbf{J}_{p}
\end{eqnarray}
where $\hat{\kappa}$ is the unit vector along the $z$ axis and the subscripts 1 and 2 refer to the two half spaces separated by the crystal. Without loss of generality I restrict the treatment to an $s$ polarized incident wave. 

\begin{figure}
\includegraphics{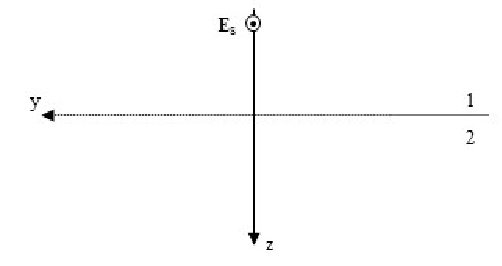}
\caption{Electric field of an $s$ polarized plane wave normally incident on a single-layer two-dimensional crystal. The two half spaces separated by the atomically thin crystal are indicated as 1 and 2.}
\end{figure}

From equations (\ref{current density}-\ref{Boundary}), I obtain:  
\begin{eqnarray}
\label{Fresnel}
E_{xi}+E_{xr}&=&E_{xt} \\
H_{yi}-H_{yr}&=&H_{yt}+\dot{P}= H_{yt}-\frac{2}{\eta}E_{R}\nonumber
\end{eqnarray}
where the subscripts $i, r, t$ denote the incident, the reflected and the transmitted waves. Defining $r_{s}=E_{r}/E_{i}$ and $t_{s}=E_{t}/E_{i}$ as the reflection and the transmission coefficients, for the square and the triangular lattice I have
\begin{eqnarray}
\label{Fresnel1}
r_{s}=\frac{E_R}{E_i}=-\frac{iN\alpha k}{2(1-\frac{C_0\alpha}{4\pi a^3})+iN\alpha k}; \qquad t_{s}=r_{s}+1
\end{eqnarray}
where eq. (39) of ref. \cite{Luca16} has been used. For the honeycomb lattice using eqs. (43) and (44) of ref. \cite{Luca16} I obtain
\begin{eqnarray}
\label{Fresnel2}
r_{s}=-\frac{A}{2B+A}; \qquad t_{s}=r_{s}+1
\end{eqnarray}
Where
\begin{eqnarray}
\label{Fresnel3}
A&=&ikN\bigg(\alpha_1+\alpha_2-\frac{2\alpha_1\alpha_2\left(C_0^{(1)}-C_0^{(2)}\right)}{4\pi a^3}\bigg) \\
B&=&1-\frac{C_0^{(1)}(\alpha_1+\alpha_2)}{4\pi a^3}+\frac{\alpha_1\alpha_2\left(C_0^{(1)2}-C_0^{(2)2}\right)}{(4\pi a^3)^2}
\end{eqnarray}

\section{Connection between the radiation-reaction electric field and the macroscopic field}

I can now establish an important connection in between $\vec{\textbf{\emph{E}}}_{R}(t)$ and the macroscopic electric field $\vec{\textbf{\emph{E}}}(t)$. From formula (\ref{Fresnel1}) it is clear that for an $s$ polarized incident wave 
\begin{eqnarray}
\label{Reaction_reflection2}
\vec{\textbf{\emph{E}}}_{R}(t) =r_s \vec{\textbf{\emph{E}}}_{i}(t)
\end{eqnarray}
Formula (\ref{Reaction_reflection2}) states that $\vec{\textbf{\emph{E}}}_{R}(t)$ is the macroscopic electric field irradiated by the 2D crystal. It gives rise to the reflected field and in superposition with $\vec{\textbf{\emph{E}}}_{i}(t)$ to the transmitted field i.e to the total macroscopic electric field \cite{Merano16, Luca16}. Algebraically, this may be written as
\begin{eqnarray}
\label{Reaction_Macro}
\vec{\textbf{\emph{E}}}(t) =\vec{\textbf{\emph{E}}}_{R}(t)+\vec{\textbf{\emph{E}}}_{i}(t)
\end{eqnarray}
This field is continuous and well-defined also at the crystal location.

It is now possible to deduce the expression of $\chi$ as a function of $\alpha$ using both the local field computed with the retarded potentials in ref. \cite{Luca16} and the radiation-reaction electric field
\begin{eqnarray}
\label{chi}
\chi = \frac{\vec{\textbf{\emph{P}}}(t)}{\epsilon_{0}  \vec{\textbf{\emph{E}}}(t)}= \frac{N\alpha \epsilon_0\vec{\textbf{\emph{E}}}_{loc}(t)}{ \epsilon_{0} ( \vec{\textbf{\emph{E}}}_ i(t)+ \vec{\textbf{\emph{E}}}_ R(t))}
\end{eqnarray}
As an example I compute $\chi$ for the square and the triangular lattice. From equations (\ref{Local1}), (\ref{Reaction_square}), and eq. (39) of \cite{Luca16}
\begin{eqnarray}
\label{chi_square}
\chi &=& \frac{N\alpha \epsilon_0\vec{\textbf{\emph{E}}}_{loc}}{ \epsilon_{0} \big(\vec{\textbf{\emph{E}}}_{loc}
\left(1-\frac{\alpha C_0}{4\pi a^3}-i \frac{\alpha C_1 k}{4\pi a^2}\right)+\frac{i\, \alpha C_1 k}{4\pi a^2} \vec{\textbf{\emph{E}}}_{loc}\big)} \nonumber \\
&=&\frac{N\alpha}{1-\frac{C_0\alpha}{4\pi a^3}}
\end{eqnarray}
It is now possible to write in a more compact form the reflection coefficient. From formula (\ref{Fresnel1})
\begin{eqnarray}
\label{Fresnel4}
r_s=-\frac{ik\chi}{2+ik\chi}
\end{eqnarray}
It is easy to verify that the same expression for the reflection coefficient is valid also for the honeycomb lattice. As expected the Fresnel coefficients depend on the atomic polarizability $\alpha$ only through the macroscopic surface susceptibility $\chi$. Starting from a complete microscopic theory the connection to the macroscopic fields has been achieved. The validity of these expressions has already been proven in experiments \cite{Merano16}. 

\section{Magnitude and phase of the radiation-reaction electric field}

In this section a numerical example is provided to demonstrate the impact of the radiation-reaction electric field. Formulas (\ref{Reaction_square}) and  (\ref{Reaction_honey}) demonstrate that $\vec{\textbf{\emph{E}}}_{R}$ is dephased from the local electric field by $\pi/2$. I consider an insulating 2D crystal with a typical $\chi =10^{-9}$ m \cite{Merano16, Merano316} at a wavelength of 633 nm. From formula (\ref{Reaction_reflection2}) it is clear that $\vec{\textbf{\emph{E}}}_{R}$ is dephased from $\vec{\textbf{\emph{E}}}_{i}$ of $\pi/2 -k\chi /2 \approx \pi/2 - 1/200$ and that its magnitude is $ \approx k\chi /2 = 1/200$ that of $\vec{\textbf{\emph{E}}}_{i}$. From formula (\ref{Reaction_reflection2}), $\vec{\textbf{\emph{E}}}_{R}$ has the same magnitude and phase of the reflected field. This shows that for an $s$ polarized plane wave normally incident on an insulating 2D crystal, the reflected field is dephased by almost $\pi/2$; in contrast to a very thin dielectric slab (erroneously used to fit the optical response of a single-layer 2D crystal \cite{Merano16}) or a bulk semi-infinite dielectric where the phase difference is $\pi$. An analogous argument holds also for $p$ polarized light.

\section{Work done by the radiation-reaction force}

For the square and the triangular lattice, the time-average work ($W_R$) per unit time and per unit area done by the radiation-reaction electric field is given by $N$ times the work done on a dipole
\begin{eqnarray}
\label{time-average work}
W_R&=&\frac{N}{T}\int_{0}^{T} Re\bigg( \vec{\textbf{\emph{E}}}_{R}(t)\bigg) \cdot Re\bigg( \dot{\vec{\textbf{\emph{p}}}}(t)\bigg) dt=-\frac{R\vec{\textbf{\emph{E}}}^2_{i}}{\eta}
\end{eqnarray} 
where $R$ is the reflectivity of the crystal \cite{Merano16}. For a honeycomb lattice the time-average work per unit time and per unit area done by the radiation-reaction electric field is given by $N$ times the one done on a primitive cell 
\begin{eqnarray}
\label{time-average work honeycomb}
W_R&=&\frac{N}{T}\int_{0}^{T} Re\bigg( \vec{\textbf{\emph{E}}}_{R}(t)\bigg) \cdot Re\bigg( \dot{\vec{\textbf{\emph{p}}}}_1 (t)+\dot{\vec{\textbf{\emph{p}}}}_2 (t)\bigg) dt =-\frac{R\vec{\textbf{\emph{E}}}^2_{i}}{\eta}
\end{eqnarray} 
The right-hand sides of equations (\ref{time-average work}) and (\ref{time-average work honeycomb}) are negative and equal in absolute value to the time-average intensity radiated by the 2D crystal \cite{Jackson}. Half of this electromagnetic intensity has an opposite direction to the incident field and it is the reflected field (Fig. 1). The other half travels in the same direction of the incident field and it recombines with it to create the transmitted electromagnetic field. Since I am considering an insulating crystal the  time-average work per unit time and unit area done by the radiation-reaction electric field is opposite to the work done by the incident field
\begin{eqnarray}
\label{time-average work incident}
W_i&=&\frac{N}{T}\int_{0}^{T} Re\bigg( \vec{\textbf{\emph{E}}}_{i}(t)\bigg) \cdot Re\bigg( \dot{\vec{\textbf{\emph{p}}}}(t)\bigg) dt
=\frac{R\vec{\textbf{\emph{E}}}^2_{i}}{\eta}
\end{eqnarray} 

\section{Conclusions}

An insulating 2D crystal has been modelled as a 2D Bravais lattice of dipoles. If an electromagnetic plane wave is incident on the crystal, each dipole will scatter it. The effect of the dipoles is obtained by superposition of the scattered wavelets. The scattered radiation will combine coherently to generate the macroscopic transmitted and reflected fields.

The total electric field for this system is the linear superposition of the incident electric field plus the retarded field generated by all the dipoles. The local electric field acting on a single dipole is the linear superposition of the incident electric field plus the retarded fields generated by all the other dipoles. The macroscopic electric field entering in the boundary conditions for the system is provided by the linear superposition of the incident electric field plus the radiation-reaction electric field. This last result is the main contribution of this article. 

Wheeler and Feynman, in a very nice paper on the mechanism of the radiative reaction, conclude that radiation is a phenomenon as much of statistical mechanics as of pure electrodynamics \cite{Wheeler45}. Here a fundamental statistical aspect of the radiation-reaction electric field for a 2D crystal has been enlightened. A relation in between the microscopic and the macroscopic description has been found. Remarkably the advanced fields play an important role in this context. This theory is based on a well-defined and relativistically invariant prescription \cite{Dirac38, Wheeler45}, and it is well supported by experiments \cite{Merano16, Blake2007, Nair2008, Kravets2010}. Starting from a microscopic theory it connects the microscopic and the macroscopic electric fields in the crystal via the radiation-reaction electric field and the boundary conditions for the system.

The approach derived here is completely original if compared to others usually used to treat the optical response of 2D systems like metasurfaces \cite{Kuester03}. Traditional theories in metasurfaces require the calculation of the field of a small disk, which is then subtracted from the macroscopic electric field, to provide the local electric field \cite{Kuester03}. The results of the present paper are strictly valid for single-layer 2D crystals. Anyway the developed method \cite{Luca16} paves the way to a complete treatment of the optical response of few-layer atomic crystals, hetero-structures \cite{Geim14} and metasurfaces \cite{Kuester03, Capasso14, Arbabi15, Shalaev13, Brongersma14}.

\section{ACKNOWLEDGMENTS}
I acknowledge Luca Dell'Anna for useful discussions.

\bibliography{letter2}
\end{document}